\documentclass[10pt]{article} 
\usepackage{graphicx}
\usepackage{textcomp}
\usepackage{amsmath}
\usepackage{commath}
\usepackage[hmargin=1.5cm,vmargin=1.5cm]{geometry}
\usepackage[font=small,margin=0.cm,figurename=FIGURE]{caption}
\usepackage{abstract}
\usepackage{bm}
\usepackage{caption}
\usepackage{subcaption}
\usepackage{color}
\usepackage[square,numbers]{natbib}
\begin{document}
	
	\title{Wavenumber-space band clipping in nonlinear periodic structures} 
	
	\author{Weijian Jiao\footnote{jiaox085@umn.edu} \ and Stefano Gonella\footnote{sgonella@umn.edu}}
	
	\date{{\normalsize Department of Civil, Environmental, and Geo- Engineering\\University of Minnesota, Minneapolis, MN 55455, USA}\\}	

	\maketitle
	\begin{abstract}
		
		\small{In weakly nonlinear systems, the main effect of cubic nonlinearity on wave propagation is an amplitude-dependent correction of the dispersion relation. This phenomenon can manifest either as a frequency shift or as a wavenumber shift depending on whether the excitation is prescribed as a initial condition or as a boundary condition, respectively. Several models have been proposed to capture the frequency shifts observed when the system is subjected to harmonic initial excitations. However, these models are not compatible with harmonic boundary excitations, which represent the conditions encountered in most practical applications. To overcome this limitation, we present a multiple scales framework to analytically capture the wavenumber shift experienced by dispersion relation of nonlinear monatomic chains under harmonic boundary excitations. We demonstrate that the wavenumber shifts result in an unusual dispersion correction effect, which we term wavenumber-space band clipping. We then extend the framework to locally-resonant periodic structures to explore the implications of this phenomenon on bandgap tunability. We show that the tuning capability is available if the cubic nonlinearity is deployed in the internal springs supporting the resonators.}
		
		\vspace{0.4cm}
	\end{abstract}
		
\section{Introduction}
Recently, nonlinear periodic structures and metamaterials have drawn great attention due to their rich dynamic response and uniquely tunable dispersive properties that endow them with a wider functionality space compared to their linear counterparts. For example, metamaterials manufactured using soft materials can undergo large deformation that can be exploited to trigger strong nonlinear effects, including dramatic geometrical transformations \cite{Yohai_2020} and instabilities \cite{Wang_2014}.  Similar nonlinear effects can also be achieved through Hertzian contact interactions in granular crystals \cite{Daraio_2006, Boechler_2010} and magnetic interactions in magneto-mechanical systems \cite{Bilal_2017_PNAS}. 
A special attribute of nonlinear systems is the tunability of their dynamic properties, which can overcome the inherent passivity of linear phononic crystals and metamaterials (a review of these effects is given by \cite{Bertoldi_2017}). Nonlinearity can also be exploited to enable solitary wave propagation in soft metamaterials \cite{Raney_PNAS_2016, Deng_2017_PRL, ZIV_2020}.

In weakly nonlinear systems, a well-known phenomenon associated with quadratic nonlinearity is the second harmonic generation (SHG). While SHG has been widely studied and observed in conventional solids and structures \cite{polyakova_1964, Hamilton_2003,Matlack_2011}; and an excellent review can be found in \cite{Hamilton}, its implications for periodic structures have only come to prominence in recent years. Among the early contributions in this direction, we recall the work on SHG in monatomic chains \cite{Tournat_2013, Mehrem_2017}. A number of works have presented applications of this (or similar) nonlinear phenomenon germane to the metamaterial paradigm, including modal mixing \cite{Ganesh_2017}, acoustic diodes and switches \cite{Liang_2010, Boechler_2011}, and subwavelength energy harvesting \cite{JIAO2018, Jiao_2018_PRA}. 

Other weakly nonlinear systems of interest are those featuring weak cubic nonlinearity. The main effect of cubic nonlinearity on wave propagation is an amplitude-dependent correction of the dispersion relation, which can manifest either as a frequency shift or as a wavenumber shift depending on the variable that is controlled in the excitation \cite{Jiao_2019_PRE}. For example, if the excitation is prescribed as an initial condition (i.e., an initial spatial profile prescribed over the domain), one can consider the wavenumber as a fixed externally-controlled parameter. In this case, the cubic nonlinear effect manifests as a frequency shift. In contrast, working with a boundary excitation implies that the input frequency can be treated as the fixed parameter, and therefore the effect has to manifest as a wavenumber shift. Various perturbation techniques have been employed to predict the dispersion shifts experienced by harmonic waves in cubic nonlinear chains. \citet{CHAKRABORTY_2001} proposed a perturbation scheme to determine the amplitude-dependent characteristic of the frequency cutoffs of nonlinear monatomic chains. \citet{LAZAROV_2007} employed the method of harmonic balance to study the shift of bandgaps in chains with attached nonlinear resonators. Later, \citet{Narisetti_2010} derived an explicit expression for frequency shifts using the \textit{Lindstedt-Poincar\'{e}} perturbation technique. A more general analytical treatment based on multiples scales analysis can be found in textbooks on perturbation methods \cite{Holmes}, leading to a consistent result for frequency shifts if harmonic initial excitations are prescribed. In addition, these analytical tools have been applied to higher-dimensional and multi-degree of freedom periodic structures to capture the spectro-spatial effects of the tuning of their dispersive properties \cite{Narisetti_2011, MANKTELOW_2013, Zhou_2018, Bukhari_2020, CAMPANA_2020}.

While the bulk of the literature on the subject is focused on frequency shifts because of the mathematical tractability of the problem, the theoretical framework cannot be automatically transported to the dual scenarios in which the excitation is prescribed as harmonic oscillations at the boundaries. This condition is of greater interest for practical engineering applications, in which the excitation is indeed prescribed using a point source through an actuation device, such as a shaker or transducer. Adapting an framework previously introduced to study doubly-nonlinear systems that feature simultaneously quadratic and cubic nonlinearities \cite{Jiao_2019_PRE}, in this paper we use a multiple scales scheme to properly determine the dispersion correction of weakly nonlinear chains under boundary excitations (i.e., wavenumber shifts). In Section 2, we demonstrate that cubic nonlinearity, in combination with harmonic boundary excitations, gives rise to unusual dispersive properties that are fundamentally different from those associated with frequency shifts, and all these findings are supported by numerical simulations. In Section 3, the framework is extended to locally-resonant periodic structures, in which the additional degree of freedom induced by the internal resonator provides opportunities to explore the potential of wavenumber shifts for bandgap tunability under practical excitation constrains. 

\section{Dispersion relation of nonlinear monatomic chain}
In this section, we develop a multiple scales framework to properly capture the dispersion relation of a nonlinear monatomic spring-mass chain under both initial and boundary excitations, and we investigate how different excitation conditions can significantly change the manifestation of cubic nonlinearity on the dispersive properties. 

\subsection{Multiple scales analysis}
Consider an infinite monatomic chain in which adjacent masses are connected by springs featuring cubic nonlinearity. Under the assumption of weak nonlinearity, the restoring force in each spring can be expressed as 
\begin{equation}\label{Restoring force}
f=k\delta+\epsilon \Gamma \delta^3 
\end{equation}
where $\delta$ denotes the relative displacement, $\epsilon$ is a small parameter, and $k$ and $\epsilon\Gamma$ are the linear and cubic spring constants, respectively. The equation of motion for the $n^{th}$ mass $m$ can be derived as
\begin{equation}\label{governing_eqns}
m \ddot{u}_n + k (2u_n-u_{n-1} -u_{n+1}) + \epsilon \Gamma \left[ (u_n-u_{n-1})^3 - (u_{n+1}-u_n)^3\right]    = 0
\end{equation}
where $u_n$ represents the displacement of the $n^{th}$ mass and the superscripted dot denotes time differentiation.

In the spirit of multiple scales analysis, we introduce a fast spatio-temporal variable $\theta_n=\xi n-\omega t$ (where $\xi$ and $\omega$ are the nondimensional wavenumber and frequency, respectively) to capture the fundamental wave response, and two slow variables $s=\epsilon n$ (spatial), and $\tau=\epsilon t$ (temporal) to capture the weakly nonlinear effects (i.e., dispersion corrections). Accordingly, the solution is assumed to have an expansion of the form
\begin{equation}\label{soln_assumed}
u_n \approx  u^0_n(\theta_n,s,\tau) +\epsilon u^1_n(\theta_n,s,\tau) + O(\epsilon^2)
\end{equation}
Substituting Eq.~\ref{soln_assumed} into Eq.~\ref{governing_eqns} and expanding $u_{n-1}$ and $u_{n+1}$ into Taylor series with respect to $s$, we obtain the equations at each order of expansion:
\begin{align}
\begin{split}\label{order1}
O(1)&:\quad \omega^2 m \frac{\partial^2 u^0_n}{\partial \theta_n^2}+k (2u^0_n-u^0_{n-1} -u^0_{n+1})=0
\end{split}
\\
\begin{split}\label{order2}
O(\epsilon)&:\quad \omega^2 m \frac{\partial^2 u^1_n}{\partial \theta_n^2}+k (2u^1_n-u^1_{n-1} -u^1_{n+1})=f
\end{split}
\end{align}
where the forcing function $f$ at $O(\epsilon)$ is given as
\begin{equation}\label{nforcef}
\begin{split}
f=2\omega m\frac{\partial^2 u^0_n}{\partial \theta_n \partial \tau}+k\left(\frac{\partial u^0_{n+1}}{\partial s}- \frac{\partial u^0_{n-1}}{\partial s} \right) - \Gamma\left[  (u^0_n-u^0_{n-1})^3 - (u^0_{n+1}-u^0_n)^3 \right] 
\end{split}
\end{equation}
The general solution at $O(1)$ can be expressed as 
\begin{equation} \label{soln1_O1}
u_n^0=\frac{A(s,\tau)}{2} e^{i\theta_n}+\frac{A^*(s,\tau)}{2}  e^{-i\theta_n}
\end{equation}
where $A$ is an arbitrary function of the slow variables $(s,\tau)$, and $(\cdot)^*$ denotes the complex conjugate. Imposing Bloch conditions on the fast scale variable $\theta_n$ between neighboring masses, we obtain
\begin{equation}\label{soln2_O1}
u^0_{n\pm 1}=\frac{A(s,\tau)}{2}  e^{i\theta_n}e^{\pm i\xi}+\frac{A^*(s,\tau)}{2} e^{-i\theta_n}e^{\mp i\xi}
\end{equation}
Substituting Eq.~\ref{soln1_O1} and Eq.~\ref{soln2_O1} into Eq.~\ref{order1}, the linear dispersion relation is obtained 
\begin{equation}\label{eigenvaule_prob}
\omega=\sqrt{2k(1-\cos(\xi))/m}
\end{equation}

\begin{figure} [!htb]
	\centering
	\includegraphics[scale=0.65]{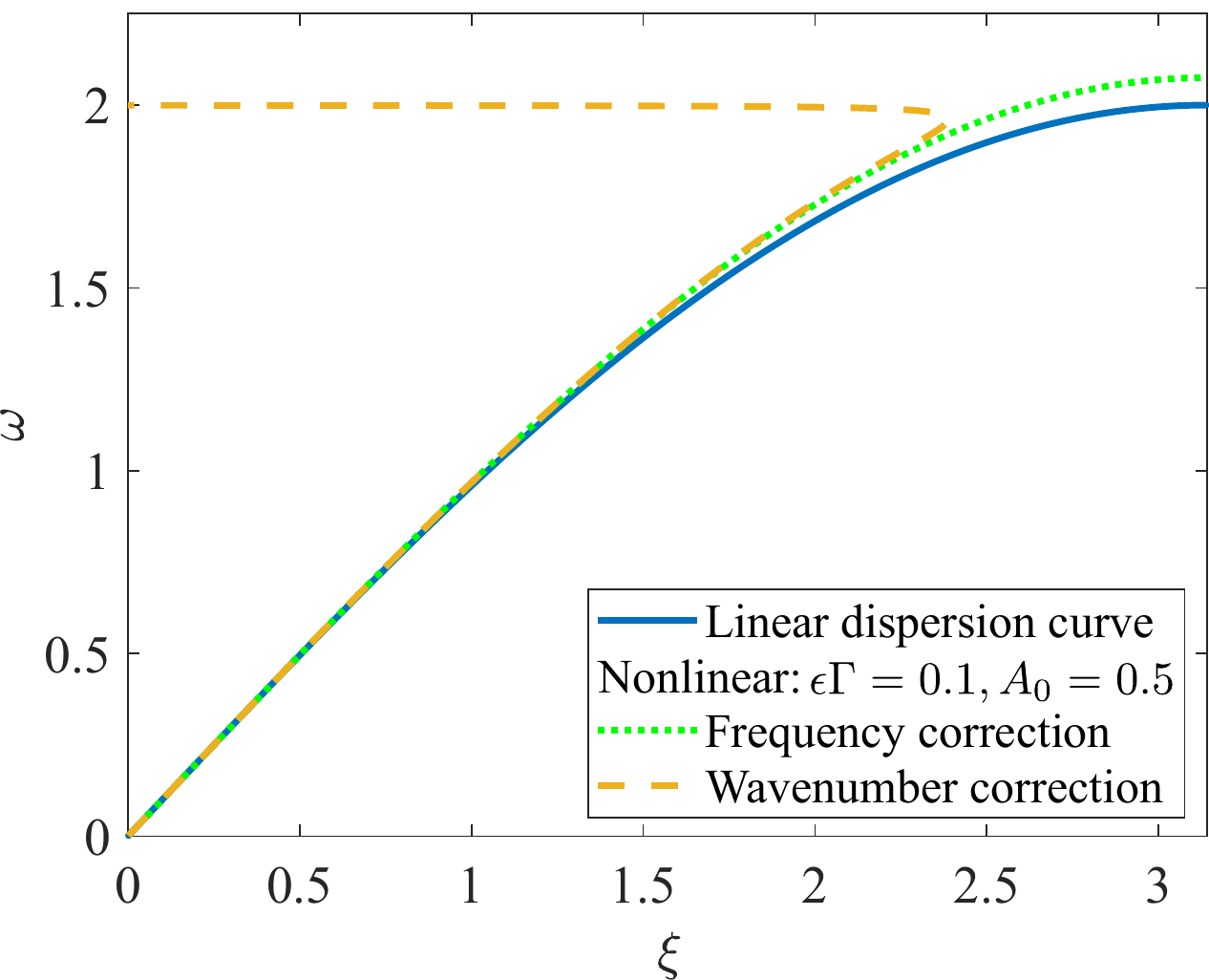}
	\caption{Nonlinear dispersion relation of a nonlinear monatomic chain obtained through standard multiple scales analysis (divergence occurs in the case of wavenumber correction).}
	\label{Band_NMC_old}
\end{figure}

\begin{figure} [!htb]
	\centering
	\includegraphics[scale=0.65]{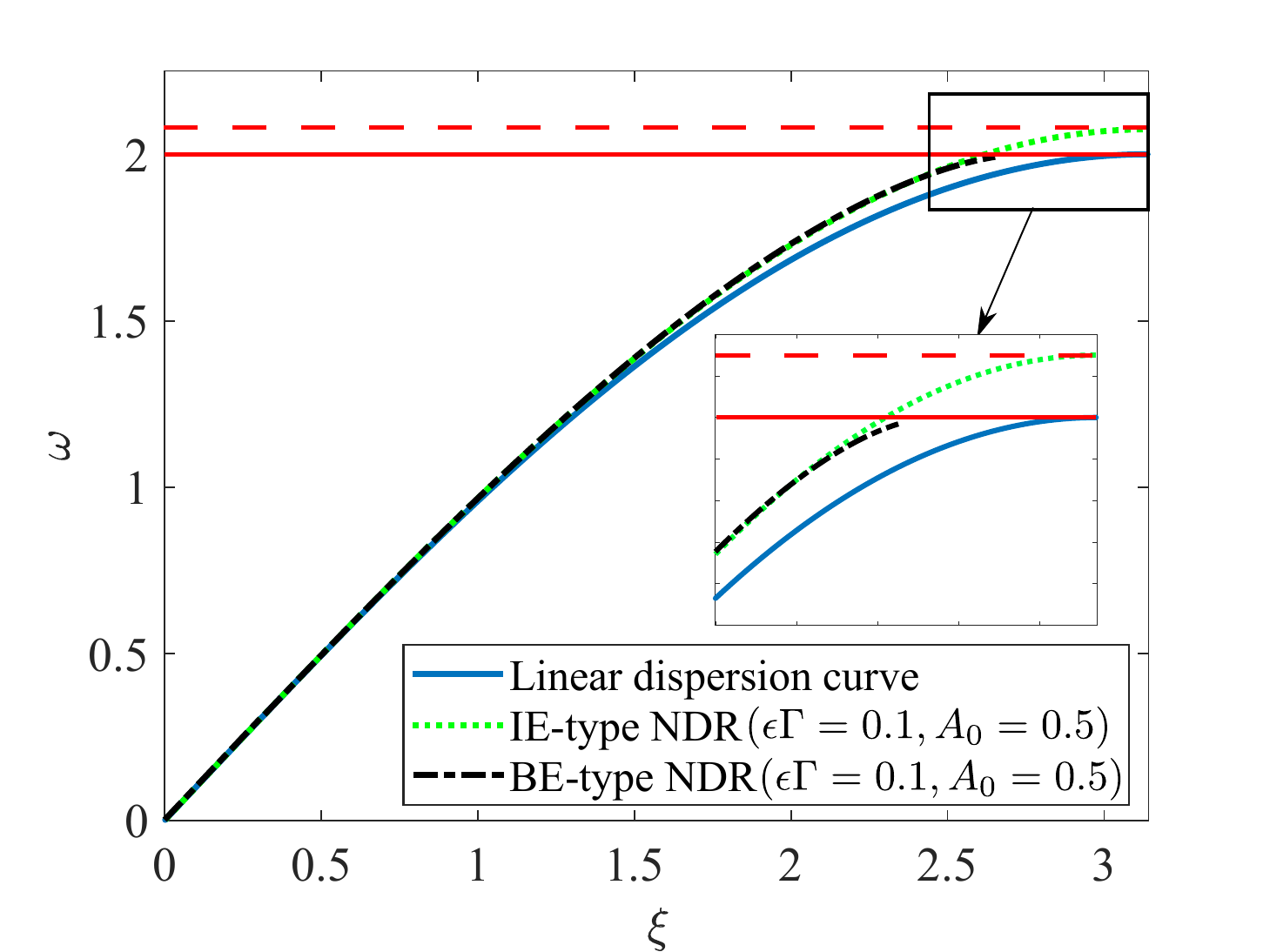}
	\caption{Corrected dispersion relation of a monatomic chain featuring hardening cubic nonlinearity obtained for different types of excitation conditions (a detailed comparison near the cutoff frequency is given in the inset). Red solid line: cutoff frequency of BE-type NDR (identical to the linear counterpart). Red dashed line: cutoff frequency of IE-type NDR.}
	\label{Band_NMC_new}
\end{figure}
We now shift our attention to the equation at $O(\epsilon_1)$, in which the forcing function $f$ can be obtained by substituting Eq.~\ref{soln1_O1} and Eq.~\ref{soln2_O1} in Eq.~\ref{nforcef}. To prevent unbounded solution, the secular terms (i.e., the terms of the form $e^{\pm i\theta_n}$) appearing in $f$ must be eliminated, leading to the following condition
\begin{equation}\label{secularity_eqn}
\frac{\partial A}{\partial \tau}+\lambda \frac{\partial A}{\partial s}+i\mu \abs{A}^2 A=0
\end{equation}
where $\lambda=k \sin \xi /\omega m$ and $\mu=6\Gamma \sin^4(\xi/2)/\omega m$. The solution of $A$ can be written in a polar form as
\begin{equation}\label{Amplitude}
A\left(s,\tau \right) = \alpha\left(s,\tau \right)e^{-i\beta\left(s,\tau \right)}
\end{equation}
Substituting it into Eq.~\ref{secularity_eqn}, yields a complex-variable algebraic equation, the solution of which requires that real and imaginary components vanish individually such that 
\begin{align}\label{Amplitude_eqns}
\begin{split}
\frac{\partial \alpha}{\partial \tau}+\lambda \frac{\partial \alpha}{\partial s}&=0
\\
\frac{\partial \beta}{\partial \tau}+\lambda \frac{\partial \beta}{\partial s}&=\mu \alpha^2
\end{split}
\end{align}
The general solutions of these equations are 
 \begin{align}\label{Amplitude_soln}
 \begin{split}
\alpha&=\alpha_0\left(s-\lambda \tau \right) \\
\beta&=\beta_0\left(s-\lambda \tau \right) + \beta^*
 \end{split}
 \end{align}
where both $\alpha_0$ and $\beta_0$ are functions of $s-\lambda \tau$ that can be determined using excitation conditions. The full expression for $\beta$ encompasses a homogeneous solution $\beta_0$ and a particular solution $\beta^*$, the latter of which can be expressed either in terms of variable $\tau$ as $\beta^*=\mu \alpha^2 \tau$, or in terms of variable $s$ as $\beta^*=\mu \alpha^2 s/\lambda$, depending on whether initial conditions or boundary conditions are considered. For example, given the initial harmonic amplitude profile $u_n=A_0 \cos \xi n$ at $t=0$, it follows that $\alpha=A_0$, $\beta_0=0$ and $\beta^*=\mu \alpha^2 \tau$. Thus, the fundamental solution at $O(1)$ is
\begin{equation}\label{modified_soln1_O1}
u_n^0=\frac{A_0}{2} e^{i\left[ \xi n-\left(\omega + \epsilon \mu A^2_0 \right) t\right] }+c.c.
\end{equation}
where $c.c.$ denotes the complex conjugate of the preceding term.  Eq.~\ref{modified_soln1_O1} shows that the nonlinear equation of motion (i.e., Eq.~\ref{governing_eqns}) produces a fundamental plane-wave solution, in which the frequency is modified by an amplitude-dependent correction term $\epsilon \mu A^2_0$. This cubic nonlinear effect is well documented in mathematics textbooks \cite{Holmes}, as well as in literature of nonlinear periodic structures \cite{Narisetti_2010}. In contrast, if a boundary condition $u_n=A_0 \cos \omega t$ is imposed at one end of the chain (e.g., at $n=0$), it follows that $\alpha=A_0$, $\beta_0=0$ and $\beta^*=\mu \alpha^2 s/\lambda$. With this, the fundamental solution becomes
\begin{equation}\label{modified_soln2_O1}
	u_n^0=\frac{A_0}{2} e^{i\left[ \left( \xi-\epsilon \mu A^2_0/\lambda \right) n-\omega t\right] }+c.c.
\end{equation}
in which the correction term $\epsilon \mu A^2_0/\lambda$ takes place in the wavenumber domain.

Clearly, either frequency shifts or wavenumber shifts can modify the dispersive properties of the nonlinear monatomic chain. To demonstrate their tuning effects on the dispersion relation, in Fig.~\ref{Band_NMC_old} we plot the modified dispersion relations predicted by Eq.~\ref{modified_soln1_O1} and Eq.~\ref{modified_soln2_O1}, in comparison with the linear one, using the following parameters:  $m=1$, $k=\Gamma=1$, $\epsilon=0.1$, $A_0=0.5$ (standard SI units are adopted throughout the article and omitted for simplicity). From a visual inspection, we observe that the two nonlinear dispersion branches start to deviate from the linear curve as $\xi$ increases. While the two nonlinear curves overlap for a large range of frequencies, the one predicted by Eq.~\ref{modified_soln2_O1} diverges when $\omega$ is close enough to the cutoff frequency (2 rad/s). This divergence issue implies that the weakly nonlinear assumptions do not hold any more, and therefore the predictions of Eq.~\ref{modified_soln2_O1} are spurious.

To resolve this issue, we follow the approach proposed by \citet{Jiao_2019_PRE}, the basic notion of which is that the Bloch condition (Eq.~\ref{soln2_O1}), which imposes a constrain on the wavenumber between neighboring masses, needs to be updated once a wavenumber correction is determined from the perturbation analysis. Specifically, the original wavenumber $\xi$ should be replaced by the modified one $\xi-\epsilon \beta/s$ (this can be obtained by substituting Eq.~\ref{Amplitude} in Eq.~\ref{soln1_O1} and collecting all the wavenumber contributions). This treatment results in a new condition (replacing Eq.~\ref{secularity_eqn}) for the elimination of the secular terms at $O(\epsilon)$:
\begin{equation}\label{secularity_eqn_new}
\frac{\partial A}{\partial \tau}+\lambda_0 \sin \left( \xi-\epsilon\frac{\beta}{s}\right) \frac{\partial A}{\partial s}+i\mu \abs{A}^2 A=0
\end{equation}
where $\lambda_0=\frac{k }{\omega m}$. 
Substituting Eq.~\ref{Amplitude} in Eq.~\ref{secularity_eqn_new} yields the following equations for $\alpha$ and $\beta$
\begin{equation}\label{secularity_eqn_new2}
\begin{split}
\frac{\partial \alpha}{\partial \tau}+\lambda_0 \sin\left( \xi-\epsilon\frac{\beta}{s}\right) \frac{\partial \alpha}{\partial s}&=0
\\
\frac{\partial \beta}{\partial \tau}+\lambda_0 \sin\left( \xi-\epsilon\frac{\beta}{s}\right)  \frac{\partial \beta}{\partial s}&=\mu \alpha^2
\end{split}
\end{equation}
It can be shown that the above system of equations reduces to Eq.~\ref{Amplitude_eqns} to the first-order approximation for cases where the wavenumber correction $\epsilon\beta/s$ is at a higher order when compared to $\xi$. However, it is possible that, for some special cases, Eq.~\ref{secularity_eqn_new2} gives profoundly different results from those solved by Eq.~\ref{Amplitude_eqns} as demonstrated below. Considering the same boundary excitation condition, it is reasonable to assume a priori, if a plane wave solution is allowed, that $\alpha=A_0$, $\beta_0=0$ and $\beta^*=C s=\epsilon Cn$, where $C$ is a real constant. Here, $\epsilon C$ is the ``true'' wavenumber shift that we intend to determine. Substituting these into Eq.~\ref{secularity_eqn_new2}, yields the following transcendental equation for $C$
\begin{equation}\label{secularity_eqn_new3}
\lambda_0 C \sin \tilde{\xi}=\mu A_0^2
\end{equation}
where $\tilde{\xi}=\xi- \epsilon C$. While Eq.~\ref{secularity_eqn_new3} is not amenable for analytical treatments, it is not difficult to find possible solutions numerically. Once $C$ is determined, the fundamental solution can be expressed as
\begin{equation}\label{modified_soln_new}
u_n^0=\frac{A_0}{2} e^{i\left[ (\xi-\epsilon C) n-\omega t\right]  }+c.c
\end{equation}
 For the same parameters used above, in Fig.~\ref{Band_NMC_new} we plot the corrected dispersion relation according to Eq.~\ref{modified_soln_new}, and we also superimpose the linear dispersion curve and the nonlinear one predicted by Eq.~\ref{modified_soln1_O1}, for comparison. We notice that the divergence observed in Fig.~\ref{Band_NMC_old} for frequencies close to 2 rad/s is successfully resolved. The two nonlinear dispersion curves shown in Fig.~\ref{Band_NMC_new} predict distinct dispersive characteristics near the cutoff frequency. Specifically, the cubic nonlinear effect manifests as frequency shifts when initial excitations are imposed, leading to an extension of the dispersion relation in the frequency domain (green dotted curve). In contrast, wavenumber shifts are induced when boundary excitations are imposed. As a consequence, the dispersion relation is clipped in wavenumber space near the $\pi$ limit (black dash-dotted curve). This wavenumber-space clipping effect is qualitatively consistent with observations recently reported by \citet{Bae_2020} invoking different modeling arguments. However, as shown in the inset of Fig.~\ref{Band_NMC_new}, the cutoff frequency of the nonlinear band with wavenumber corrections remains the same as the linear counterpart (note that it is true for any $A_0$ within the weakly nonlinear limits). For convenience, we will refer to the former nonlinear dispersion relation as IE-type NDR (IE standing for ``initial excitation'') and to the latter one as BE-type NDR (BE standing for``boundary excitation'') throughout this article.
 
 \begin{figure} [!htb]
 	\centering
 	\includegraphics[scale=0.65]{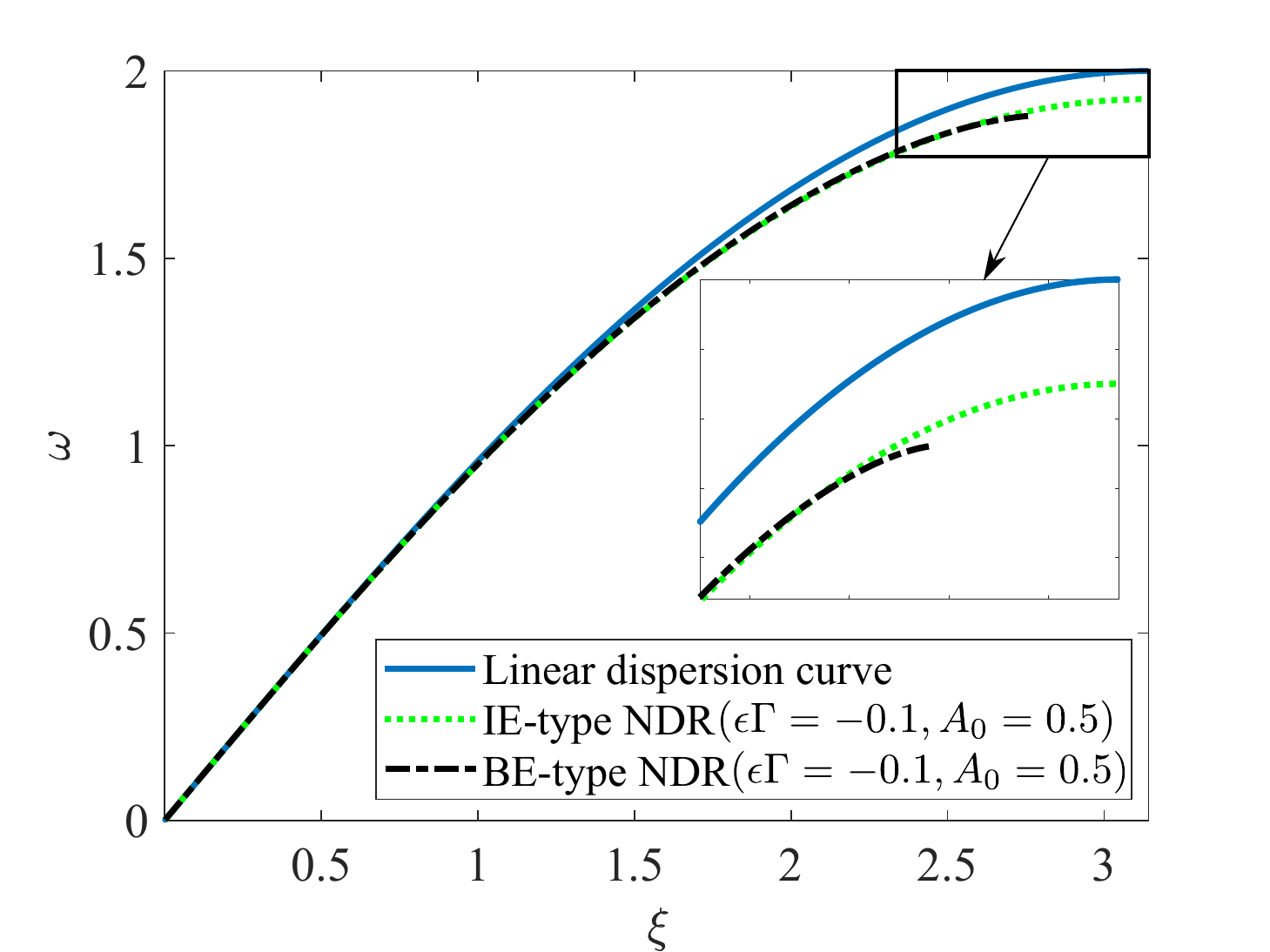}
 	\caption{Corrected dispersion relation of a monatomic chain featuring softening cubic nonlinearity obtained for different types of excitation conditions (a detailed comparison near the cutoff frequency is given in the inset).}
 	\label{Band_NMC_new2}
 \end{figure}

 In the previous example, we investigated the dispersion characteristics of a monatomic chain featuring hardening cubic nonlinearity with $\Gamma=1$. We now proceed to examine the case with softening cubic nonlinearity ($\Gamma=-1$). Following the same analysis, we can obtain the BE-type NDR of the chain with $\Gamma=-1$, which is plotted as a black dashed curve in Fig.~\ref{Band_NMC_new2} and compared against the IE-type NDR counterpart (green dotted curve). Again, we observe the band clipping effect near the $\pi$ limit. In contrast with the previous case, the cutoff frequency of the chain is modified using both nonlinear corrections. Compared to the IE-type NDR, the BE-type NDR further lowers the cutoff frequency for the same excitation amplitude, thus enabling a lager degree of nonlinear tuning.
 
 \begin{figure*} [!htb]
 	\centering
 	\includegraphics[scale=0.4]{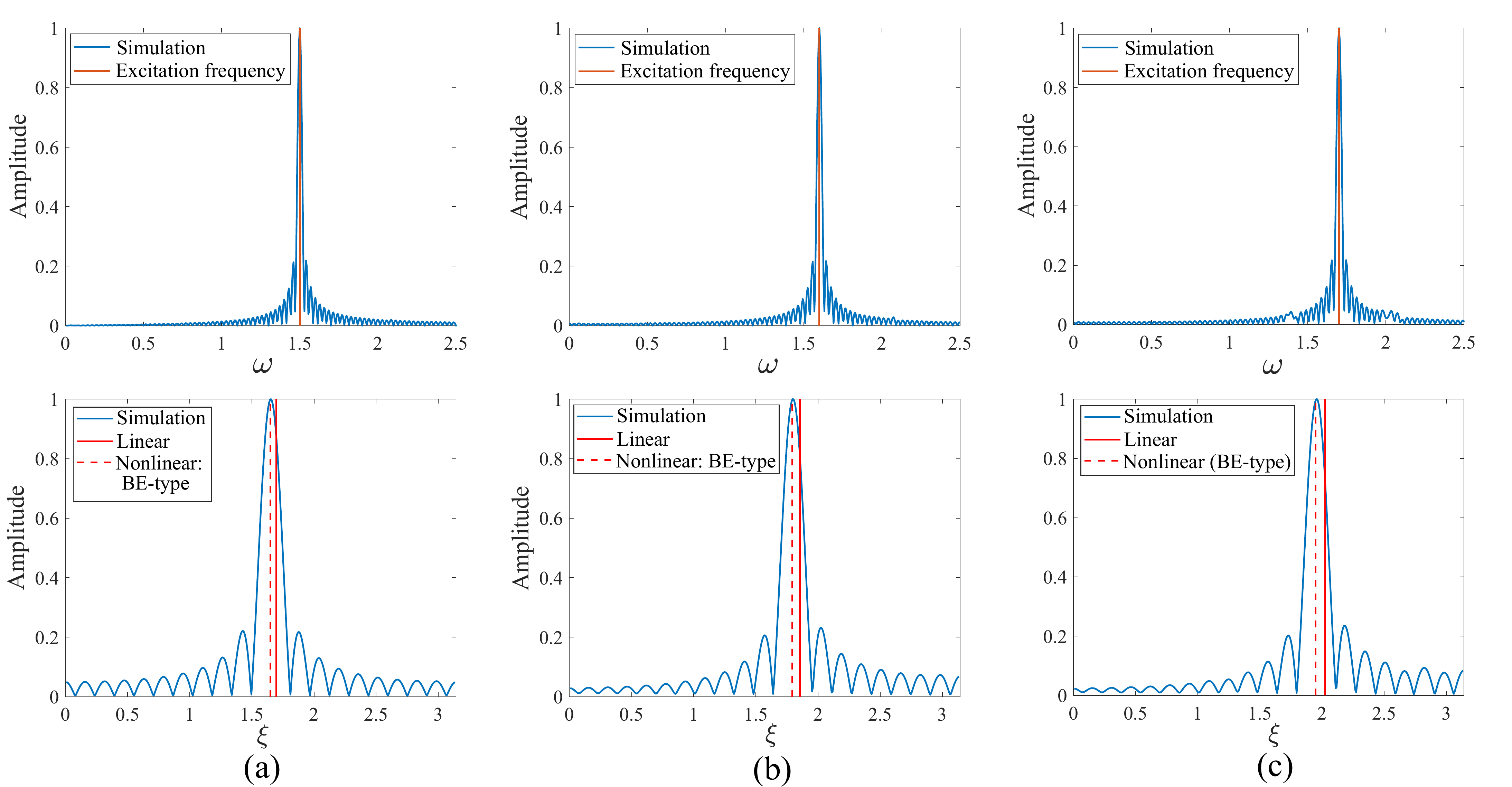}
 	\caption{Normalized spectra of the nonlinear response of a monatomic chain featuring hardening cubic nonlinearity ($\Gamma=1$) under harmonic boundary excitations with three different frequencies: (a) 1.5 rad/s; (b) 1.6 rad/s; (c) 1.7 rad/s. First row: Normalized FFT of the output signal. Second row: Normalized FFT of the spatial profile after appropriately long simulation time.}
 	\label{FFTs_monatomic}
 \end{figure*}
 
 \begin{figure*} [!htb]
 	\centering
 	\includegraphics[scale=0.6]{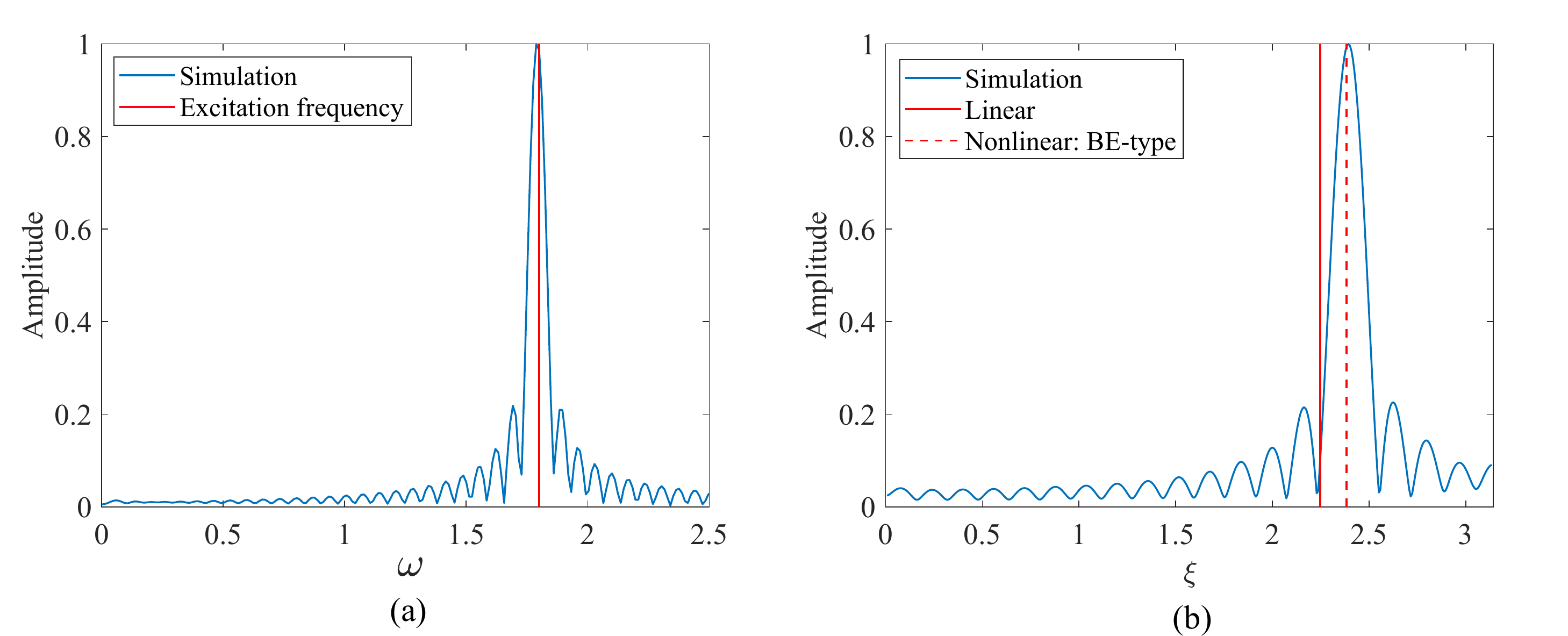}
 	\caption{Normalized spectra of the nonlinear response of a monatomic chain featuring softening cubic nonlinearity ($\Gamma=-1$) under harmonic boundary excitations at 1.8 rad/s. (a) Normalized FFT of the output signal. (b) Normalized FFT of the spatial profile after appropriately long simulation time. }
 	\label{FFTs_monatomic_Neg}
 \end{figure*}

\subsection{Full-scale simulation}

To validate the above theoretical findings, we perform numerical simulations by integrating the equations of motion (Eq.~\ref{governing_eqns}) by means of the Verlet Algorithm \cite{Swope_1982}. We study the wave response of a monatomic chain with hardening cubic nonlinearity (with $\Gamma=1$), consisting of 40 masses, and subjected to a harmonic boundary excitation imposed at the first mass. To minimize reflections and establish plane waves, an absorbing layer of 30 masses is attached at the end of the monatomic chain. In Fig.~\ref{FFTs_monatomic}, we plot the normalized FFT of the time history of the last mass (i.e., the output of the chain) and the normalized FFT of the spatial profile (after the initial transient response is damped out and nearly plane wave response is reached) for three prescribed frequencies (denoted by the red solid bars): 1.5 rad/s, 1.6 rad/s, and 1.7 rad/s. We observe that the excitation frequencies are preserved in the output time histories while wavenumber shifts are observed in the nonlinear spatial profiles, matching the prediction from Eq.~\ref{modified_soln_new} (red dashed bars) and confirming the statement that, for harmonic boundary excitations, cubic nonlinear effect manifests as wavenumber shifts, as opposed to frequency shifts. For completeness, we repeat the simulation for a chain with softening cubic nonlinearity ($\Gamma=-1$) at one excitation frequency (1.8 rad/s), and the corresponding result is given in Fig.~\ref{FFTs_monatomic_Neg}. It can be seen that the simulation result reveals a right-shifted wavenumber, which is in excellent agreement with the theoretical predication (red dashed bar). As an interesting side observation, we notice that the numerical simulation could diverge from the analytical model and multi-frequency quasi-periodic response would emerge if the harmonic excitation is set sufficiently close to the cutoff frequency and beyond certain amplitude thresholds (similar phenomena are reported by \citet{Boechler_2011}).

\begin{figure} [!htb]
	\centering
	\includegraphics[scale=0.6]{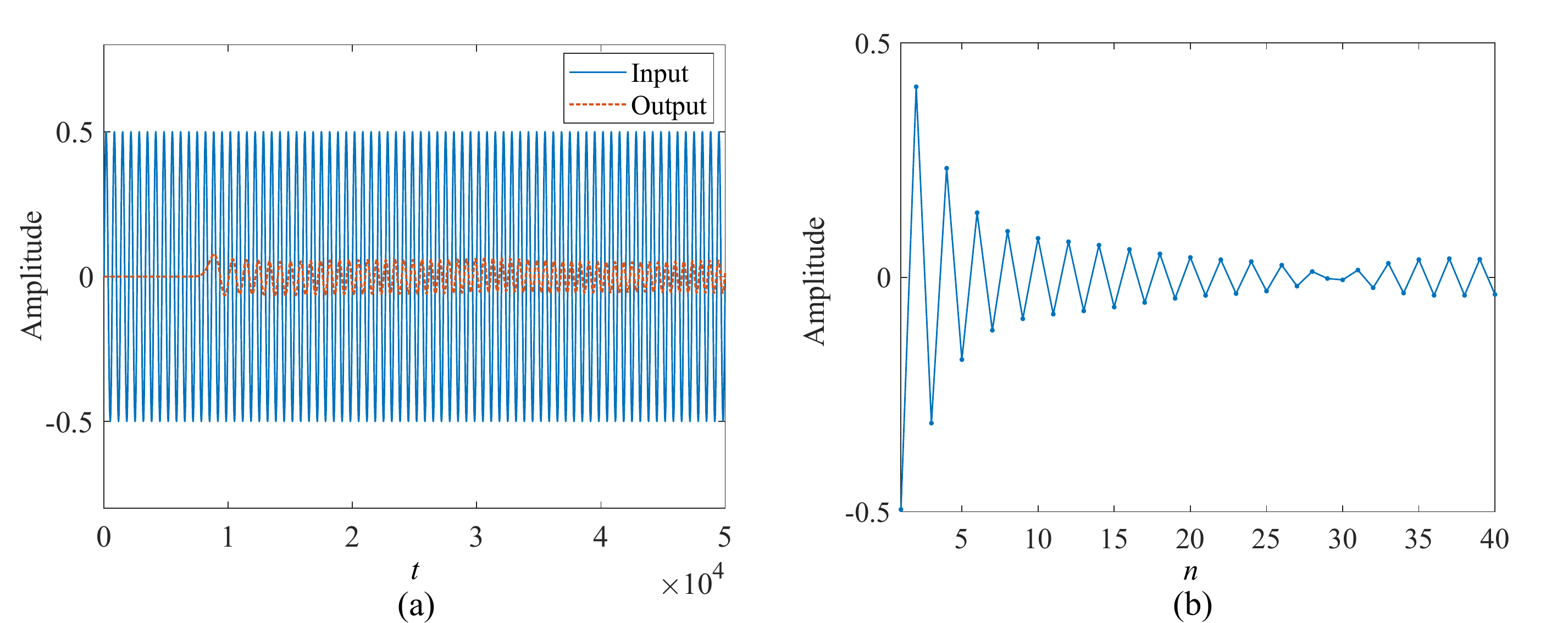}
	\caption{Nonlinear response of a monatomic chain under harmonic boundary excitation at frequency 2.05 rad/s, lying above the cutoff of the linear system but below that of the branch endowed with frequency correction. (a) Input signal vs. output signal. (b) Spatial profile after appropriately long simulation time. Both data sets suggest attenuation compatible with bandgap conditions, confirming that a frequency shift of the dispersion branch is not observed for boundary excitations.}
	\label{Attentuation_Simulation}
\end{figure}

The next task is to verify that the BE-type NDR preserves the cutoff frequency of the linear case. A characteristic that makes it distinct from the IE-type NDR for cases with hardening cubic nonlinearity. To this end, we set the excitation frequency to 2.05 rad/s, which is above the linear cutoff frequency (2 rad/s) but below that of the IE-type NDR (2.075 rad/s). The corresponding nonlinear response is shown in Fig.~\ref{Attentuation_Simulation}, in which we plot the input and output as functions of time in Fig.~\ref{Attentuation_Simulation}(a), as well as the spatial profile in Fig.~\ref{Attentuation_Simulation}(b). From a visual inspection, we observe a low output-input ratio and a clear spatial attenuation, which indicate the establishment of bandgap conditions further confirming the validity of the BE-type NDR obtained from the analytical model.

\section{Bandgap tunability in nonlinear locally-resonant periodic structures}
Metamaterials featuring internal resonators are of great interest for their ability to open locally-resonant bandgap at low frequencies \cite{Liu_2000}, as well as their implications for wave manipulation, including negative refraction \cite{Huang_2014}, subwavelength wave steering \cite{Celli_2015}, and seismic shielding \cite{Colombi_2016}. In this section, we extend the multiple scales framework to nonlinear locally-resonant periodic structures to explore the availability of bandgap tunability under the practical constrain of boundary excitations. 

\subsection{Multiple scales analysis}

\begin{figure} [!htb]
	\centering
	\includegraphics[scale=0.35]{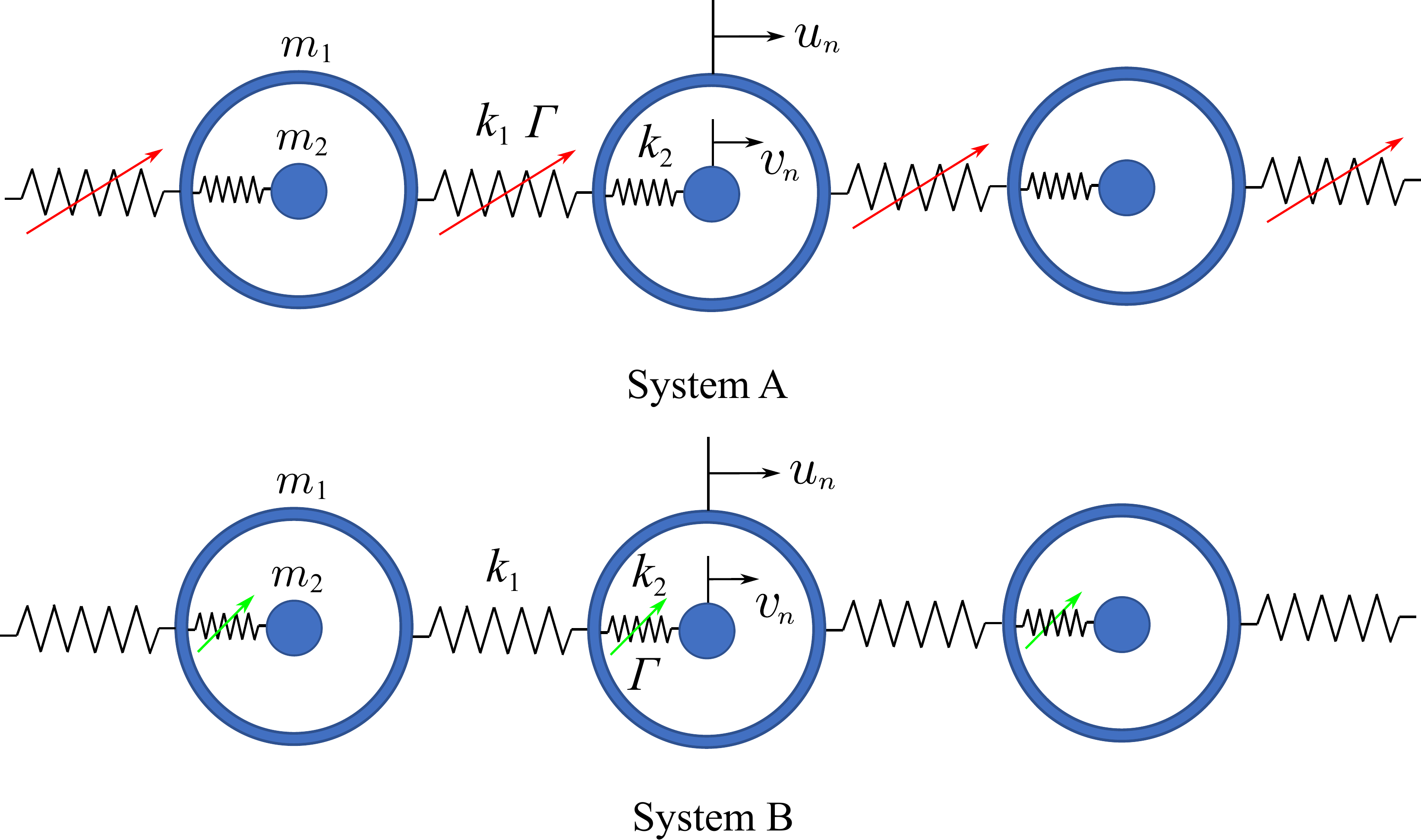}
	\caption{Schematic of nonlinear mass-in-mass chains. System A: a mass-in-mass chain with nonlinear springs connecting the masses $m_1$.  System B: a mass-in-mass chain with nonlinear springs attached to the local resonators $m_2$.}
	\label{Mass_in_Mass_Sketch}
\end{figure}

Consider a periodic structure featuring internal resonators, which can be conceptually modeled as a mass-in-mass chain (as depicted in Fig.~\ref{Mass_in_Mass_Sketch}). Cubic nonlinearity can be incorporated either in the springs connecting the neighboring masses of the main chain (configuration referred to as system A), or in the internal springs attached to the internal resonators (system B). To obtain the BE-type NDR of the two systems, we extend the multiple scales analysis of the monatomic chain to the two-degree-of-freedom problem of  the mass-in-mass chain.

The equations of motion for the two nonlinear mass-in-mass systems can be written in the matrix form as 
 \begin{equation}\label{governing_eqns_Mass_in_Mass}
\begin{split}
\begin{bmatrix} m_1 & 0 \\ 0 & m_2 \end{bmatrix} \begin{Bmatrix}\ddot{u}_n \\ \ddot{v}_n\\ \end{Bmatrix} +
\begin{bmatrix} 2k_1+k_2 & -k_2 \\ -k_2 & k_2 \end{bmatrix} \begin{Bmatrix}u_n \\ v_n\\ \end{Bmatrix} +
\begin{bmatrix} -k_1 & 0 \\ 0 & 0 \end{bmatrix} \begin{Bmatrix}u_{n-1} \\ v_{n-1}\\ \end{Bmatrix} +
\begin{bmatrix} -k_1 & 0 \\ 0 & 0 \end{bmatrix} \begin{Bmatrix}u_{n+1} \\ v_{n+1}\\ \end{Bmatrix} 
+ \mathbf{f}^{\mathrm{A(B)}}_{NL} = \begin{Bmatrix}0 \\ 0\\ \end{Bmatrix}
\end{split}
\end{equation}
in which $\mathbf{f}^A_{NL}=\epsilon \Gamma \begin{Bmatrix}  (u_n-u_{n-1})^3 - (u_{n+1}-u_n)^3  \\ 0\\ \end{Bmatrix}$ for system A and $\mathbf{f}^B_{NL}=\epsilon \Gamma \begin{Bmatrix} (u_n- v_n)^3 \\ -(u_n- v_n)^3\\ \end{Bmatrix}$ for system B, and $u_n$ and $v_n$ denote the displacements of $m_1$ and $m_2$ in the $n$th unit cell, respectively. The nodal displacements can be expressed up to the first-order approximation as 
\begin{align}\label{soln_assumed_Mass_in_Mass}
\begin{split}
\mathbf{u}_n=\begin{Bmatrix}u_{n}(t) \\ v_{n}(t)\\ \end{Bmatrix}
\approx \begin{Bmatrix}u^0_n(\theta_n,s,\tau) +\epsilon u^1_n(\theta_n,s,\tau)  \\ v^0_n(\theta_n,s,\tau) +\epsilon v^1_n(\theta_n,s,\tau) \\ \end{Bmatrix}
\end{split}
\end{align}
By substituting Eq.~\ref{soln_assumed_Mass_in_Mass} into Eq.~\ref{governing_eqns_Mass_in_Mass}, we obtain the following system of cascading equations
\begin{align}
\begin{split}\label{order1_Mass_in_Mass}
O(1)&:\quad \omega^2\mathbf{M}\frac{\partial^2 \mathbf{u}^0_n}{\partial \theta_n^2}+\mathbf{K}_1\mathbf{u}_n+\mathbf{K}_2\mathbf{u}^0_{n-1}+\mathbf{K}_2\mathbf{u}^0_{n+1}=\mathbf{0}
\end{split}
\\
\begin{split}\label{order2_Mass_in_Mass}
O(\epsilon)&:\quad \omega^2\mathbf{M}\frac{\partial^2 \mathbf{u}^1_n}{\partial \theta_n^2}+\mathbf{K}_1\mathbf{u}^1_n+\mathbf{K}_2\mathbf{u}^1_{n-1}+\mathbf{K}_2\mathbf{u}^1_{n+1}=\mathbf{F}^{\mathrm{A(B)}}
\end{split}
\end{align}
where $\mathbf{M}=\begin{bmatrix} m_1 & 0 \\ 0 & m_2 \end{bmatrix}$ is the mass matrix; $\mathbf{K}_1=\begin{bmatrix} 2k_1+k_2 & -k_2 \\ -k_2 & k_2 \end{bmatrix} $; $\mathbf{K}_2=\begin{bmatrix} -k_1 & 0 \\ 0 & 0 \end{bmatrix} $ and the forcing term at $O(\epsilon)$ is
\begin{equation}\label{forcing}
\mathbf{F}^{\mathrm{A(B)}}=2\omega \mathbf{M}\frac{\partial^2 \mathbf{u}^0_n}{\partial \theta_n \partial \tau}-\mathbf{q}^{\mathrm{A(B)}}+\begin{Bmatrix} k_1\left(\frac{\partial u^0_{n+1}}{\partial s}- \frac{\partial u^0_{n-1}}{\partial s} \right)\\ 0\\ \end{Bmatrix}
\end{equation}
in which $\mathbf{q}^A= \Gamma \begin{Bmatrix}  (u^0_n-u^0_{n-1})^3 - (u^0_{n+1}-u^0_n)^3  \\ 0\\ \end{Bmatrix}$ for system A and $\mathbf{q}^B= \Gamma \begin{Bmatrix} (u^0_n- v^0_n)^3 \\ -(u^0_n- v^0_n)^3\\ \end{Bmatrix}$ for system B.

A plane wave solution at $O(1)$ can be assumed to have the form
\begin{equation} \label{soln_O1_Mass_in_Mass}
\mathbf{u}_n^0=\frac{A(s,\tau)}{2} \boldsymbol{\phi} e^{i\theta_n}+\frac{A^*(s,\tau)}{2}  \boldsymbol{\phi}^* e^{-i\theta_n}
\end{equation}
where $\boldsymbol{\phi}=\begin{Bmatrix} \phi_u \\ \phi_v\\ \end{Bmatrix}$ is a modal eigenvector. Accordingly, the Bloch condition (Eq.~\ref{soln2_O1}) becomes
\begin{equation}\label{Bloch_soln_O1_Mass_in_Mass}
\mathbf{u}_{n\pm 1}=\frac{A(s,\tau)}{2} \boldsymbol{\phi} e^{i\theta_n}e^{\pm i\xi}+\frac{A^*(s,\tau)}{2}\boldsymbol{\phi}^* e^{-i\theta_n}e^{\mp i\xi}
\end{equation}
Substituting Eq.~\ref{soln_O1_Mass_in_Mass} and Eq.~\ref{Bloch_soln_O1_Mass_in_Mass} into Eq.~\ref{order1_Mass_in_Mass} yields the following eigenvalue problem
\begin{equation}\label{eigenvaule_prob_Mass_in_Mass}
\left( -\omega^2 \mathbf{M}+\mathbf{K}(\xi)\right) \boldsymbol{\phi}=\mathbf{0}
\end{equation}
where $\mathbf{K(\xi)}=\begin{bmatrix} 2k_1\left( 1-\cos\xi\right) +k_2 & -k_2 \\ -k_2 & k_2 \end{bmatrix}$.  The linear dispersion relation of a mass-in-mass chain can be obtained by solving the above eigenvalue problem, which can be analytically expressed as 
\begin{equation}\label{Linear_dispersion}
\omega_{1,2}=\sqrt{\frac{b\mp\Delta(\xi)}{2a}}
\end{equation}
where $\Delta(\xi)=\sqrt{b^2-4ac}$, with $a=m_1m_2$, $b=\left( m_1+m_2\right) k_2+2m_2k_1\left(1-\cos \xi \right)$, and $c=2k_1k_2 \left(1-\cos \xi \right)$, and $\omega_{1}$ and $\omega_{2}$ denote the acoustic and optical branches, respectively. The corresponding normalized eigenvector is given by
\begin{equation}\label{modal_vector}
\boldsymbol{\phi}_{1,2}=\begin{Bmatrix}-\omega^2_{1,2}m_2+k_2 \\ k_2\ \end{Bmatrix}/ \sqrt{\omega^4_{1,2}m^2_2-2k_2m_2\omega^2_{1,2}+2k^2_2}
\end{equation}

\begin{figure} [!htb]
	\centering
	\includegraphics[scale=0.6]{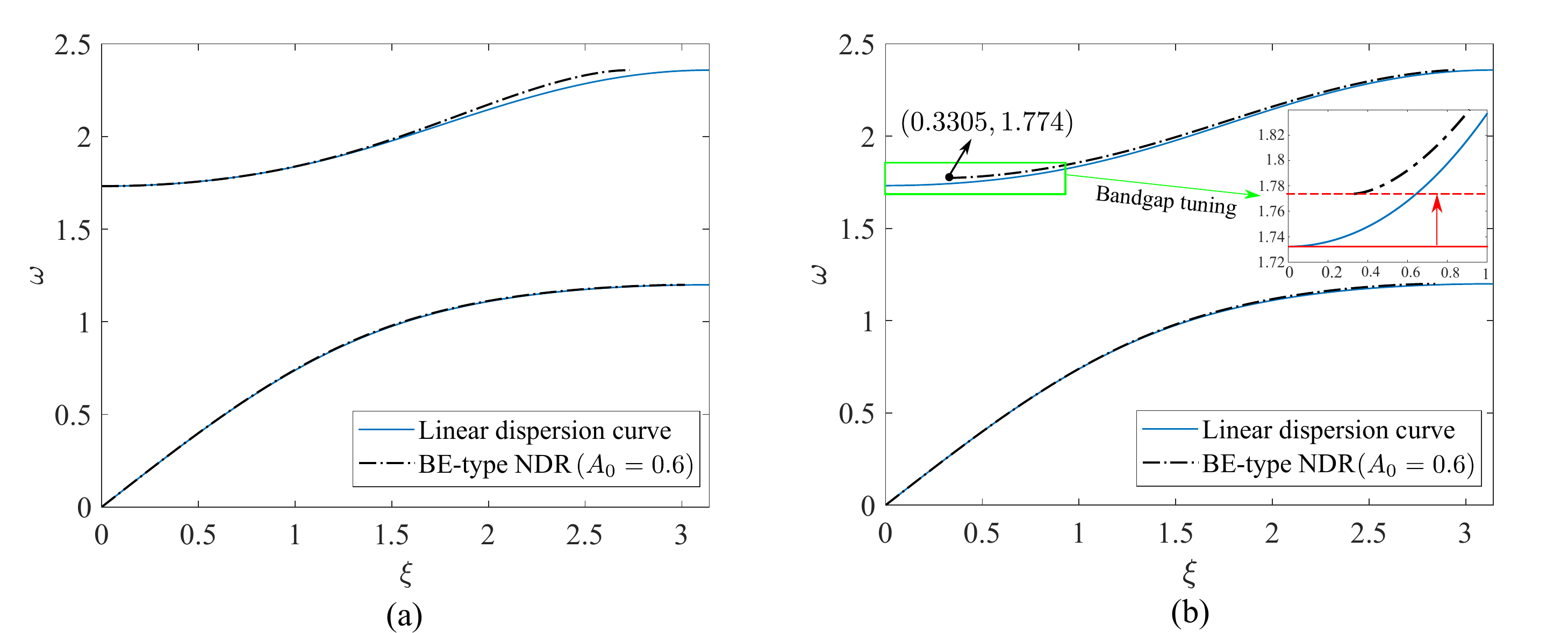}
	\caption{Corrected dispersion relations of (a) system A with nonlinearity in the main chain and (b) system B with nonlinearity in the local resonators (with values of the cut-on wavenumber and frequency of the optical branch indicated). The band clipping effect gives rise to a bandgap tuning functionality in system B, which is not available in system A.}
	\label{BE_NDR_Mass_in_Mass}
\end{figure}

We now follow the same logic described in the section 2.1 to derive an equation for the elimination of the secular terms at $O(\epsilon)$. For multi-degree-of-freedom systems, we can write the solution $\mathbf{u}^1_n$ as a superposition of the normal modes: $\mathbf{u}^1_n=\boldsymbol{\Phi} \mathbf{z}_n$, where $\boldsymbol{\Phi}$ is the modal matrix featuring the modal vectors $\boldsymbol{\phi}$ as columns \cite{Meiro, MANKTELOW_2013}. Premultiply Eq.~\ref{order2_Mass_in_Mass} by $\boldsymbol{\phi}^H$, where $(\cdot)^H$ denotes conjugate transpose, and note that its linear kernel is the same as the linear kernel of Eq.~\ref{order2_Mass_in_Mass}, leading to
\begin{equation}\label{order1_decoupled_Mass_in_Mass}
\omega^2 \bar{m}\frac{\partial^2 z_n}{\partial \theta_n^2}+\bar{k}z_n=\boldsymbol{\phi}^H \mathbf{F}^{\mathrm{A(B)}}
\end{equation}
where $\bar{m}=\boldsymbol{\phi}^H \mathbf{M}\boldsymbol{\phi}$ and $\bar{k}=\boldsymbol{\phi}^H \mathbf{K}\boldsymbol{\phi}$ are the modal mass and stiffness, respectively. The elimination of the secular terms in the RHS of Eq.~\ref{order1_decoupled_Mass_in_Mass} leads to the required condition using the updated wavenumber $\xi-\epsilon\frac{\beta}{s}$
\begin{equation}\label{secularity_eqn_new_Mass_in_Mass}
\frac{\partial A}{\partial \tau}+\lambda_0 \sin \left( \xi-\epsilon\frac{\beta}{s}\right) \frac{\partial A}{\partial s}+i\mu^{\mathrm{A(B)}} \abs{A}^2 A=0
\end{equation}
where $\lambda_0=\frac{k_1 \abs{\phi_u}^2 }{\omega \bar{m}}$, $\mu^{\mathrm{A}}=6\Gamma \phi^4_u \sin^4(\xi/2)/\omega \bar{m}$ for system A, and $\mu^{\mathrm{B}}=3\Gamma (\phi_u-\phi_v)(\phi_u^3-\phi_v^3+\phi_u \phi_v^2-\phi_u^2\phi_v)/8\omega \bar{m}$ for system B. Repeating the procedure that leads to Eq.~\ref{secularity_eqn_new3}, we derive a similar transcendental equation, from which the wavenumber shift can be determined numerically. Then, we can easily construct the BE-type NDR of the two nonlinear systems. For a set of parameters selected as $m_1=1$, $m_2=0.5$, $k_1=k_2=\Gamma=1$, $\epsilon=0.1$, and $A_0=0.5$, we plot in Fig.~\ref{BE_NDR_Mass_in_Mass} the BE-type NDR of system A in comparison with that of system B, and we superimpose their linear dispersion relations for reference. As expected, in the BE-type NDR of system A, we observe band clipping similar to that observed in the monatomic chain with a wavenumber shift that increases as $\xi$ approaches $\pi$ and no influence on the bandgap bounds. Interestingly, for system B, in addition to the $\pi$ limit, clipping also appears at the origin of the optical branch. As a result, the branch effectively starts at a finite wavenumber and the shift is accompanied by an upward shift of the cut-on frequency, resulting in a bandgap extension. This dispersive characteristic uniquely germane to system B endows it with a bandgap tuning functionality that is unachievable in system A where the nonlinearity is implemented in the main chain. 

\subsection{Full-scale simulation}

\begin{figure} [!htb]
	\centering
	\includegraphics[scale=0.4]{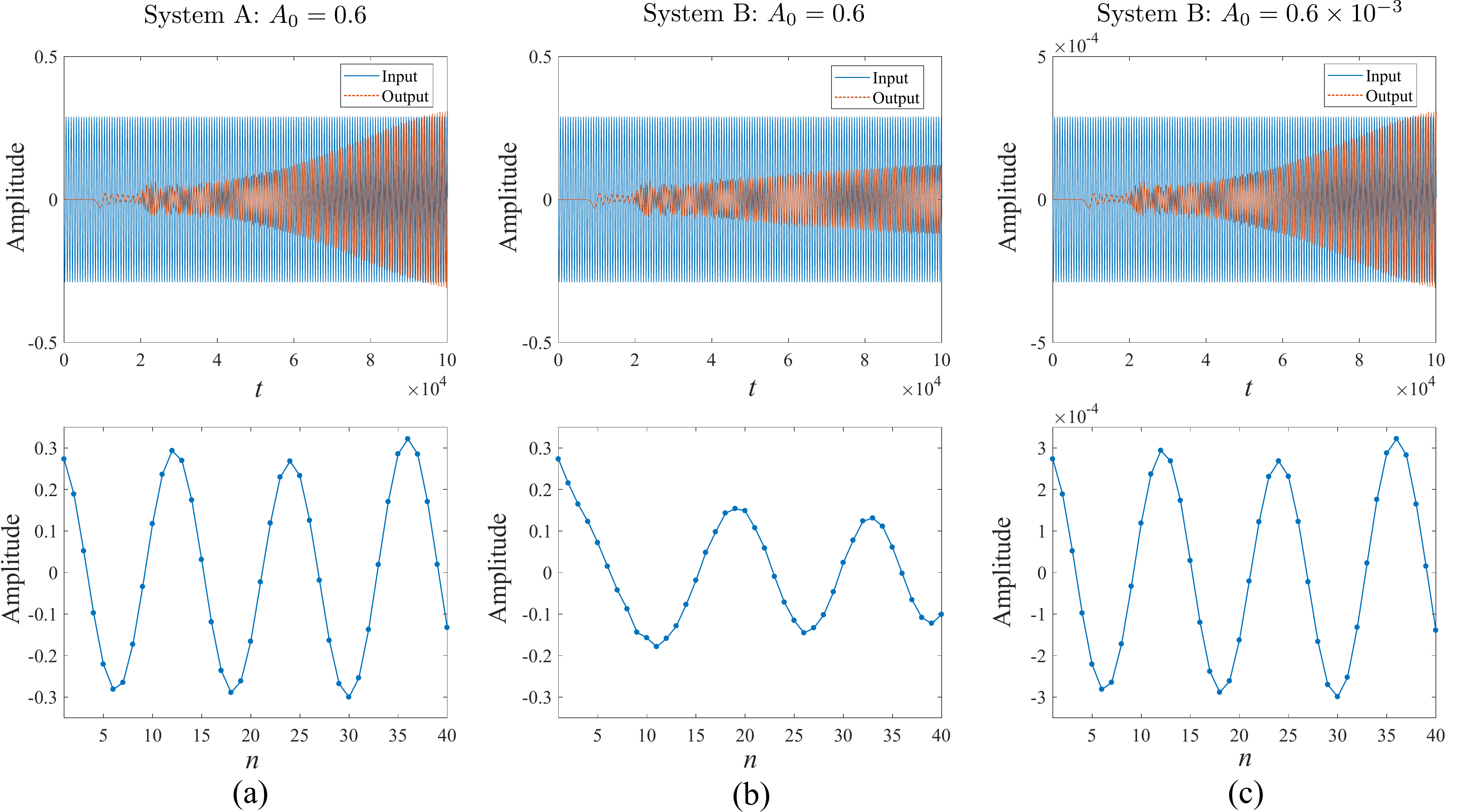}
	\caption{Temporal and spatial response of (a) system A and (b) system B for high amplitude of excitation $A_0=0.6$, and (c) system B for low amplitude $A_0=0.6\times 10^{-3}$, under harmonic boundary excitation at frequency 1.76 rad/s. First row: Input signal vs output signal. Second row: Spatial profile after appropriately long simulation time, indicating that attenuation is activated only for system B under high amplitude of excitation.}
	\label{Simulation_Mass_in_Mass}
\end{figure}

We perform a suite of numerical simulations to demonstrate the bandgap tuning capability of system B and the lack thereof of system A. The excitation frequency is set at 1.76 rad/s, which is located in the range of the bandgap extension shown in the inset of Fig.~\ref{BE_NDR_Mass_in_Mass}(b). First, in Fig.~\ref{Simulation_Mass_in_Mass}(a) and (b), we plot the temporal and spatial response of system A and B, respectively, for the same high-amplitude excitation ($A_0=0.6$). It is clear that the wave response of system B is highly attenuated compared to that of system A, confirming the existence of bandgap conditions at the prescribed frequency in system B, in accordance with the bandgap extension. Next, in Fig.~\ref{Simulation_Mass_in_Mass}(c) we plot the response of system B for a low-amplitude excitation ($A_0=0.6\times 10^{-3}$). We observe that the attenuation effect is switched off, which further demonstrates that the bandgap of system B can be tuned by controlling of the excitation amplitude. 

\begin{figure} [!htb]
	\centering
	\includegraphics[scale=0.6]{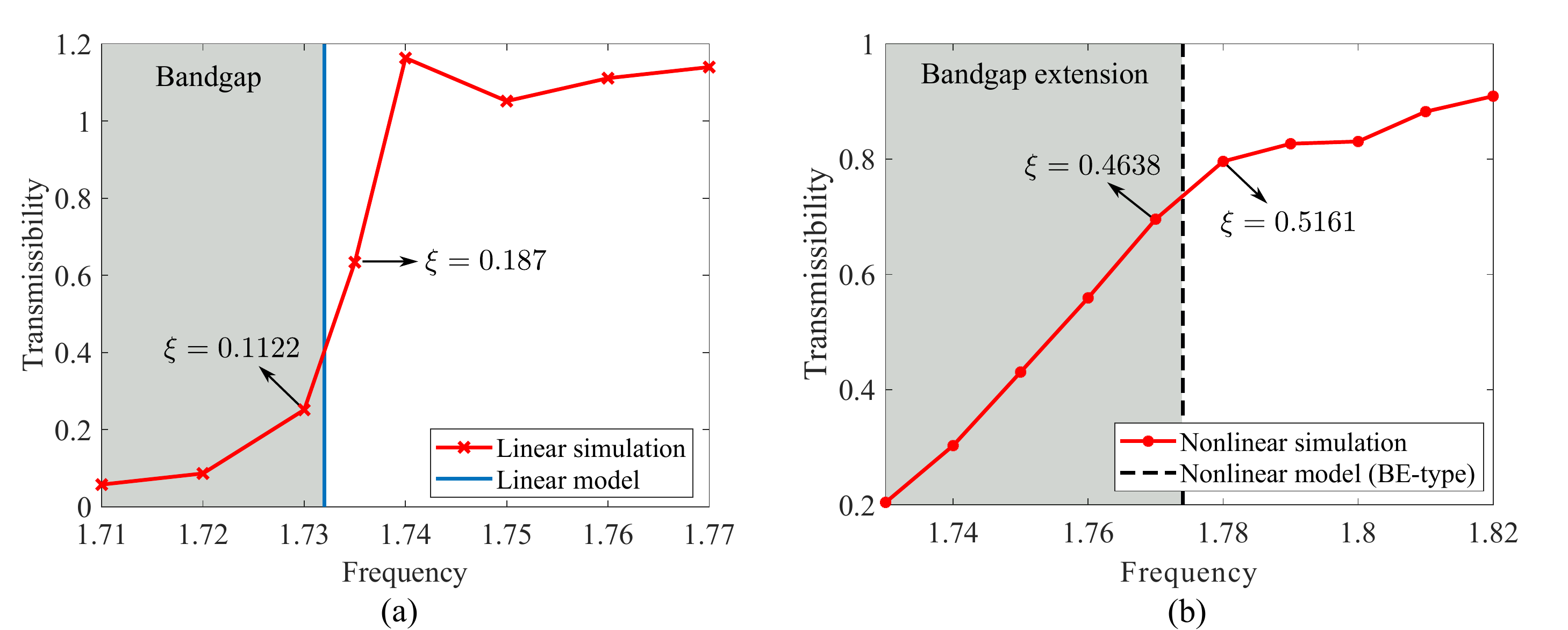}
	\caption{Transmissibility of system B around the cut-on frequency of the optical branch. (a) Linear transmissibility. (b) Nonlinear transmissibility ($\Gamma=1$, $A_0=0.6$).}
	\label{Fig2}
\end{figure}

In order to provide additional evidence that this bandgap tunability is induced by the band clipping effect, we take a closer look at the wavenumber structure of the response. In Fig.~\ref{Fig2}(a) and (b), we plot the transmissibility (red curves with markers) of system B excited in the linear and weakly nonlinear regimes, respectively. As reference, we superimpose the linear (blue solid bar in Fig.~\ref{Fig2}(a)) and the BE-type nonlinear (black dashed bar in Fig.~\ref{Fig2}(b)) cut-on frequency of the optical mode predicted by the analytical model. Moreover, for all the excitation frequencies in the considered range, we extract the wavenumber of the response by tracking the maximum value of the spectral amplitude of the spatial profile over the chain domain. In Fig.~\ref{Fig2}, we indicate the wavenumbers established for excitation frequencies across the linear and BE-type nonlinear cut-on frequency of the optical branch. In the linear case, the cut-on wavenumber falls in the range of $0.1122\sim0.187$, which can be considered close to zero. Here, the discrepancy with theory (for which the cut-on wavenumber is precisely zero) can be attributed to the impossibility to establish perfect mono-frequency plane wave conditions in the simulation. In the nonlinear case, the cut-on wavenumber jumps to finite $0.4638\sim0.5161$ range. This result confirms that the BE-type nonlinear optical mode starts at a finite wavenumber. If we subtract these numerically-obtained wavenumbers, i.e., $(0.4638\sim0.5161)-(0.1122\sim0.187)=0.3291\sim0.3516$, we obtain an approximate range of the difference of the cut-on wavenumber between the linear and nonlinear cases (incidentally, this operation can potentially cancel out the errors due to the non-idealities inevitably embedded in the simulations). Interestingly, the difference of the two cut-on wavenumbers  predicted by the analytical model (0 and 0.3305 as indicated in Fig.~\ref{BE_NDR_Mass_in_Mass}(b)) exactly falls in this range.


\section{Conclusions}
In summary, we have presented a general framework based on multiple scales analysis to properly capture the dispersive properties of weakly nonlinear periodic structures. Through this framework, we have revisited the benchmark problem of a cubic nonlinear monatomic chain, demonstrating that the cubic nonlinearity manifests as wavenumber shifts under harmonic boundary excitations in contrast to the typical frequency shifts observed when harmonic initial excitations are prescribed. Moreover, we have shown that wavenumber shifts have a strong influence on the dispersion relation, resulting in wavenumber-space band clipping. Then, we have extended the framework to nonlinear locally-resonant periodic structures to explore the bandgap tuning potential resulting from the band clipping effect. We have determined that bandgap tunability is available in systems where the cubic nonlinearity is introduced in the internal springs supporting the resonators, while no such effect arises in the presence of cubic nonlinearity in the main chain.

\section*{Acknowledgement}
The authors acknowledge the support of the National Science Foundation (CAREER Award CMMI-$1452488$). 

\bibliography{myrefs}
\end{document}